\documentclass[proof]{WileyASNA-v1}

\articletype{Article Type}%

\received{}
\revised{}
\accepted{}

\usepackage{sidecap}
\DeclareCaptionLabelFormat{continued}{#1~#2 (continued)}

\newcommand{\kms}{km\,s$^{-1}$}

\begin{document}

\title{A slowly pulsating run-away B star at high Galactic latitude ejected from a spiral arm}

\author[1]{Ulrich Heber}

\author[1]{Maximilian Halenke}

\author[2]{Aakash Bhat}

\author[3]{Veronika Schaffenroth}

\authormark{Heber \textsc{et al}}

\address[1]{\orgdiv{Dr. Remeis-Sternwarte \& ECAP}, \orgname{Friedrich-Alexander University Erlangen-N\"{u}rnberg}, \orgaddress{\state{Sternwartstr.\ 7, 96049 Bamberg}, \country{Germany}}}

\address[2]{\orgdiv{Institute for Physics and Astronomy}, \orgname{University of Potsdam}, \orgaddress{\state{Karl-Liebknecht-Str. 24/25, 
D14476 Potsdam}, \country{Germany}}}

\address[3]{\orgdiv{Thuringian State Observatory}, \orgname{} \orgaddress{\state{Sternwarte 5, D07778 Tautenburg}, \country{Germany}}}

\corres{\email{ulrich.heber@fau.de}}


\abstract{We report the discovery of the young {B6\,V} run-away star LAMOST J083323.18+430825.4, { 2.5\,kpc above} the Galactic plane.
Its atmospheric parameters and chemical composition are determined from LAMOST spectra, indicating { normal} composition. Effective temperature ($T_{\mathrm{eff}}$=$14,500$K) and gravity ($\log g$=$3.79$) suggest that the star is close to terminating hydrogen burning. An analysis of the spectral energy distribution 
allowed us to determine the angular diameter as well as the interstellar reddening. Using 
evolutionary models from the MIST database we derived the stellar mass (4.75M$_\odot$) and age 
($104^{+11}_{-13}$\,Myr). 
The spectroscopic distance { (4.17\,kpc)}, the radius (4.5\,R$_\odot$), and the luminosity ($log(L/L_\odot$)=$2.89$) then result from the atmospheric parameters. Using \textit{Gaia} proper motions, the trajectory is traced back to the Galactic disk to identify the place of birth in a spiral arm. The ejection velocity of 92\,{km\,s$^{-1}$} is typical for runaway stars in the halo. The age of the star is larger than its time of flight (78$\pm4$\,Myr), which favors a binary supernova event as the likely ejection mechanism. The \textit{TESS} light curve shows variations with a period of 3.58 days from which we conclude that it is a slowly pulsating B-star, {one} of very few run-away B-stars known to pulsate.
}

\maketitle

\section{Introduction}

Amongst the faint blue stars at high {G}alactic latitudes, young massive stars can occasionally be found far away from any star forming region. Therefore, it is likely, that they were ejected from their place of birth in the Galactic disc as runaway stars. Two ejection mechanisms have been proposed: dynamical ejection form star clusters \citep[DES,][]{runawaysblaauw2} and binary-supernova ejection \citep[BSE,][]{1967BOTT....4...86P}. In a few cases, parent star clusters have been identified \citep{hoog,bhat}
The bright (G=11.67 mag) star LAMOST J083323.18+430825.4 (J0833+4308) 
was classified as a candidate hot subdwarf star by \citet{2016MNRAS.457.3396P}. We study spectra from the LAMOST
 database and classify the star as B\,V instead. Hence, J0833+4308 is a young main sequence star at high Galactic latitude (b=36.2$^\circ$). This suggests that the star is a run-away star. We perform a comprehensive analysis combining optical spectra, the spectral energy distribution, and a \textit{TESS} light curve, with \textit{Gaia} astrometry.

\section{Spectroscopic analysis, atmospheric parameters, and chemical composition}

Two low-resolution optical spectra are available in the LAMOST database taken about 436 d apart (LAMOST J083323.17+430825.1: \#1 MJD=56973.87430556, OBSIDs: 2653010186 and \#2: MJD:57410.05546296, OBSID: 4106010186), respectively). Both spectra are of excellent quality, with a signal-to-noise ratio exceeding 100 covering the wavelength range from 3700 to 8900\AA\ which gives us access to the Balmer as well as to the Paschen series of hydrogen. A quantitative spectral analysis of those spectra was carried out using hybrid LTE/NLTE model atmospheres and synthetic spectra  
and a global fitting procedure \citep[see][for details]{andreas_spectro}. 
 
We derived an effective temperature of T$_{\mathrm{eff}}$ = 14,500\,K 
and a surface gravity of $\log (g\,\mathrm{(cm\,s^{-2})})$ = $3.79${, typical for a B6 main sequence star}.
The error budget is dominated by systematical uncertainties which we estimated to be 2\% for $T_{\mathrm{eff}}$ and 0.05 for $\log g$.
While the H and metal lines are matched very well, the cores of a few He\,{\sc i} lines (e.g. 3820, 4026, 4471\AA) are not matched quite so well. The effect is small and becomes obvious only because of the excellent SNR of the spectra.
We kept the He abundance fixed at the solar value which matches the He line profiles exempt for some cores. The projected rotation velocity turns out to be small at $<9$\,km\,s$^{-1}$ at $1\sigma$. We adopted the $30$\,km\,s$^{-1}$ ($\approx3\sigma$) as an upper limit. The radial velocities are small, 1.5 km\,s$^{-1}$ and -3.7 km\,s$^{-1}$, respectively, with formal $1\sigma$ uncertainty of $<$1 km\,s$^{-1}$, each. {The value is surprisingly small but is not exceptional among run-away B stars \citep[see][]{2011MNRAS.411.2596S}.}

Abundances of the chemical elements C, N, Mg, Si, and Fe were also derived and found to be consistent with the present-day chemical abundance standard \citep[CAS, ][]{2012A&A...539A.143N} to within 0.2 dex. Improvements to the results can be made if various correlation between projected rotational, microturbulent velocity, abundance and temperature can resolved, 
which requires high resolution spectroscopy is required. 
The results of the quantitative spectral analysis are summarized in Table \ref{tab:atmos}.

\begin{figure*}
\begin{center}
\includegraphics[width=0.9\textwidth]{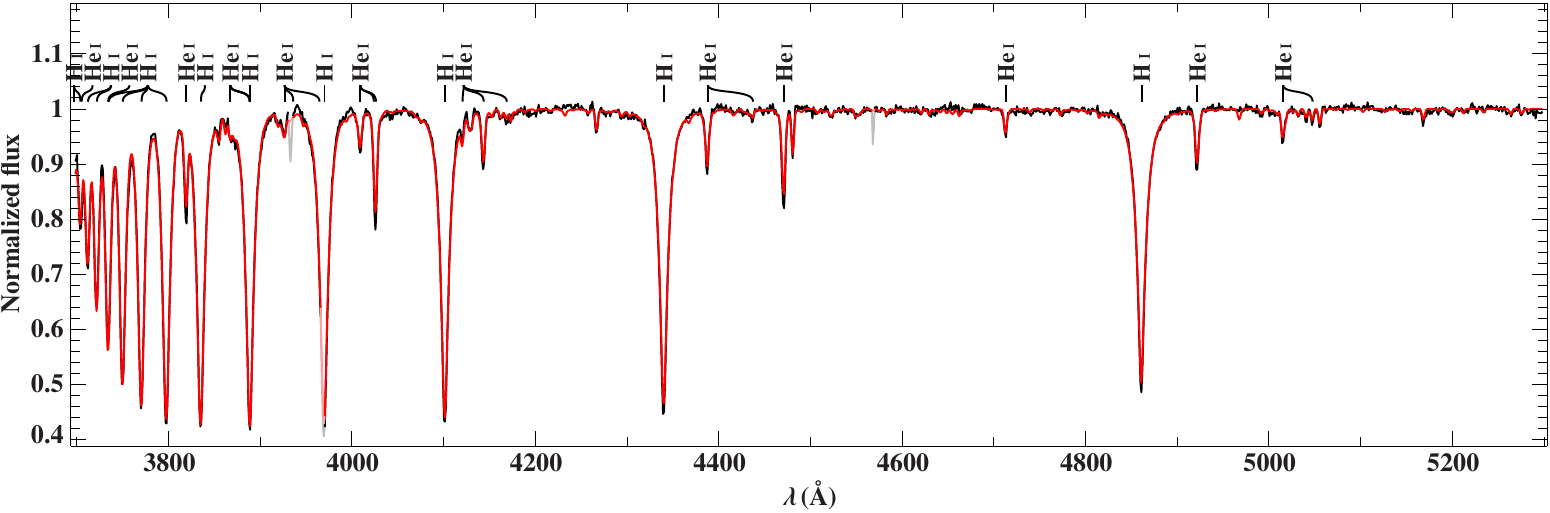}
\end{center}
    \caption{Blue part of the LAMOST spectrum \#1 of J0833+4308. 
    The best-fit synthetic spectrum (red) is shown along with the renormalised observed one. Light colours mark regions that have been excluded from fitting, e.g. the interstellar Ca\,{\sc ii} doublet.  
    }\label{fig:spectro}
\end{figure*}

Its low projected rotation velocity ($<30$\,km\,s$^{-1}$) indicates that J0833+4308 might be a slow rotator. However, it is well known, that
slowly rotating B stars
show chemical pecularities \citep{2002ApJ...573..359A}. The distribution of rotation velocities of such chemically peculiar B (Bp) stars is distinctively different from that of normal B stars, with hardly any Bp star exceeding 100\,\kms.
The fast run-away star HVS\,7, for instance, is slowly rotating ($v\sin(i)$ = 55$\pm 2$ \kms). Its gravity ($\log g$ = 3.8) is similar to that of J0833+4308. However, at $T_{\mathrm{eff}}$ =12000K,  HVS\,7 is somewhat cooler than the latter.
Its chemical abundance pattern is strikingly peculiar \citep{przybilla08}. Helium, carbon, nitrogen, and oxygen are depleted, while P and Cl as well as the iron group are enhanced by a factor of between $\approx$10 and 100, and rare-earth elements and mercury even by a factor $\approx$10 000, while manganese is undetected.
For  J0833+4308 we do not find any evidence for similar chemical pecularities, although its projected rotation velocity is similar to that of HVS\,7. This might hint at a low inclination angle of J0833+4308.

\begin{figure}
\begin{center}
\includegraphics[width=0.9\columnwidth]{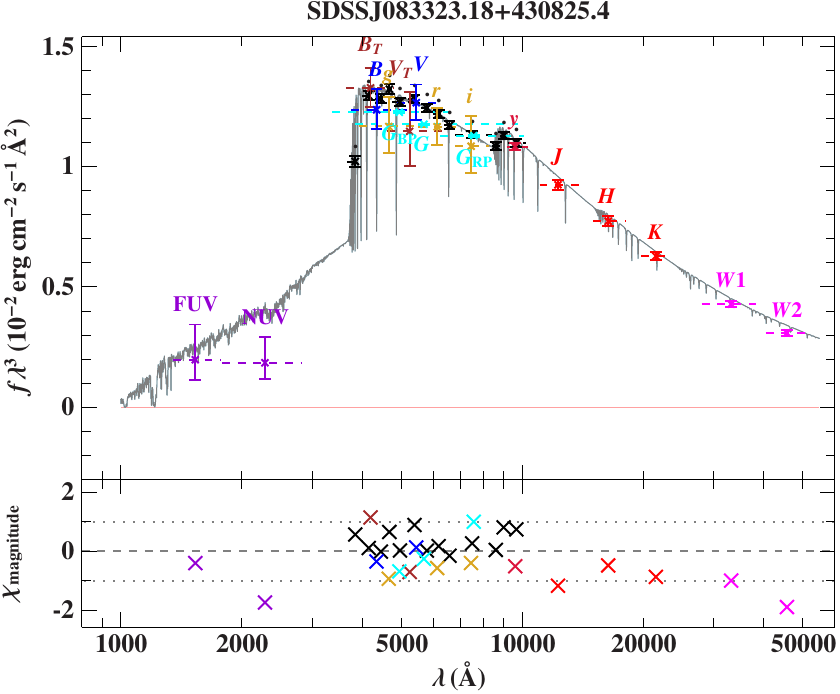}
\end{center}
\caption{Fit of the SED: Photometric fluxes are displayed as coloured data points with their respective uncertainties and filter widths (dashed lines). The best-fit models are drawn as gray full drawn lines. The lower panels gives reduce $\chi$ to demonstrate the quality of the fit.
     }
\end{figure}

\begin{table}
\renewcommand{\arraystretch}{1.2}
\caption{Atmospheric parameters and chemical abundances compared to the present-day cosmic abundance standard \citet[CAS, ][]{2012A&A...539A.143N}. The uncertainties are statistical 1$\sigma$ except for $T_{\mathrm{eff}}$ and $\log g$ (see text). }\label{tab:atmos}
\begin{tabular}{lrr}
\hline\hline
& & CAS\\
\hline
Effective temperature $T_{\mathrm{eff}}$ [K]& $14,500\pm 300$ & \\
Surface gravity $\log (g\,\mathrm{(cm\,s^{-2})})$ & $3.79\pm 0.05$ & \\
Radial velocity \#1: $v_{\mathrm{rad}}$ [km\,s$^{-1}$] & $1.9 \pm 0.7$ & \\
Radial velocity \#2: $v_{\mathrm{rad}}$ [km\,s$^{-1}$] & $-3.7 \pm 0.9$ & \\
Rotational velocity  $v\sin(i)$ [\,km\,s${}^{-1}$] & $<9$ &\\
Microturbulence $\xi$ [km\,s$^{-1}$] & $0.28^{+0.15}_{-0.22}$& \\
He abundance $\log(n(\mathrm{He}))$ (fixed) & $-1.05$ & -1.05\\
C abundance $\log(n(\mathrm{C}))$ & $-3.52\pm 0.07$ & -3.71\\
N abundance $\log(n(\mathrm{N}))$ & $-4.14^{+0.16}_{-0.19}$ & -4.25\\
O abundance $\log(n(\mathrm{O}))$ & $-3.16^{+0.05}_{-0.04}$ & -3.28\\
Mg abundance $\log(n(\mathrm{Mg}))$ & $-4.55^{+0.04}_{-0.05}$ & -4.48\\
Si abundance $\log(n(\mathrm{Si}))$ & $-4.53^{+0.06}_{-0.05}$ & -4.54\\
S abundance $\log(n(\mathrm{S}))$ & $-4.81^{+0.05}_{-0.04}$ & -4.85\\
Fe abundance $\log(n(\mathrm{Fe}))$ & $-4.47\pm 0.06$ & -4.52\\
\hline
\end{tabular}
\end{table}

\section{Spectral energy distribution} 

The spectral energy distribution (SED) has been constructed from several photometric surveys covering the UV, optical and IR part of the electromagnetic spectrum. The set of databases that have been queried is given in \citet{2024A&A...685A.134C}.
By matching synthetic SEDs to the observations \citep[see][for details]{2018OAst...27...35H} we derive an angular diameter $\log(\Theta\,\mathrm{(rad)})$ = $-10.300 \pm 0.007$ and a very small colour excess of $0.0079 \pm 0.0023$\,mag because of interstellar reddening.

\section{Kiel diagram, evolutionary age, and distance}

\begin{figure*}
\centering
\begin{minipage}{0.65\textwidth}
\centering
\includegraphics[width=0.97\textwidth]{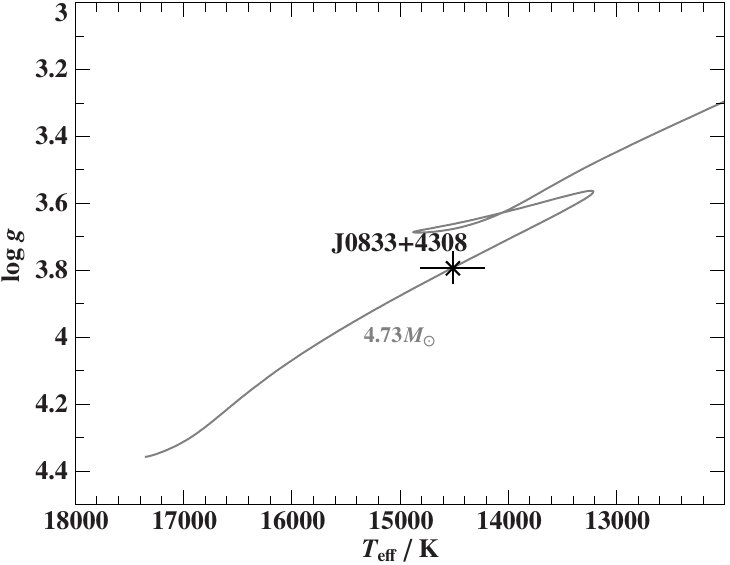}
\end{minipage}\hfill
\begin{minipage}{0.33\textwidth}
\centering
\includegraphics[width=0.99\textwidth]{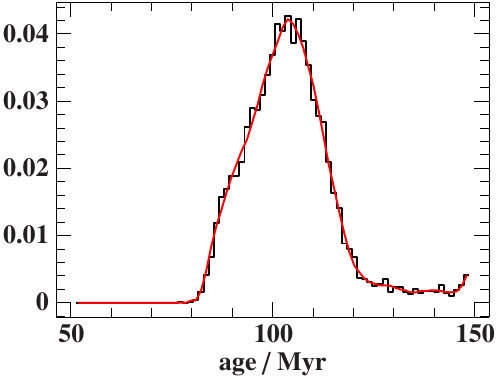}
\includegraphics[width=0.99\textwidth]{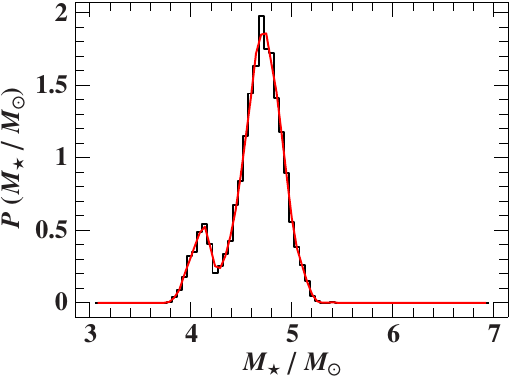}
\end{minipage}
\caption{Left: Position of J0833+4308 in the Kiel diagram with a matching evolutionary track (MIST) for a 4.8 M$_\odot$ star with solar composition. Distribution of age (middle panel) and mass (right panel) from MC simulations from MIST tracks for the atmospheric parameters of J0833+4308.}\label{fig:kiel_age_mass}. 
\end{figure*}

Comparing atmospheric parameters to stellar evolution predictions allows us to derive stellar masses as well as the age of the star (see Fig.~\ref{fig:kiel_age_mass}). Using MIST tracks \citet{choi2016} we derive a mass of M = 4.75$\pm0.24\,M_\odot$ 
and age of 104$^{+11}_{-13}$\,Myr from Monte Carlo simulations (see Fig. \ref{fig:kiel_age_mass} and Table \ref{tab:stellar}). Using the angular diameter from the SED fit, we derive a stellar radius R = $4.5^\pm0.4$\,R$_\odot$. Finally, the luminosity results from radius and effective temperature to be $\log(L/L_\odot$) = $2.89^{+0.11}_{-0.06}$. The spectroscopic distance of 4.17$\pm0.28$\,kpc is fully consistent with that derived from the \textit{Gaia} DR3
 parallax of 0.1961$\pm$0.0386\,mas \citep{2021A&A...649A...1G}, which results in 3.8$^{+0.9}_{-0.7}$,kpc , when a parallax zero point offset of -0.067 mas \citep{2021A&A...649A...2L} is added, but more precise than the latter and, therefore, preferred.

\begin{table}
\renewcommand{\arraystretch}{1.2}
\caption{Stellar parameters derived from MIST models (see Fig. \ref{fig:kiel_age_mass}). }\label{tab:stellar}
\begin{center}
\begin{tabular}{lr}
\hline\hline
mass [M$_\odot$]& $4.75\pm 0.24$\\
radius [R$_\odot$]& $4.50\pm 0.4$\\
luminosity $\log (L/L\odot)$ & $2.89^{+0.11}_{-0.06}$\\
age [Myr] & $104^{+11}_{-13}$\\
\hline
\end{tabular}
\end{center}
\end{table}

\begin{SCfigure*}
\includegraphics[width=0.55\textwidth]{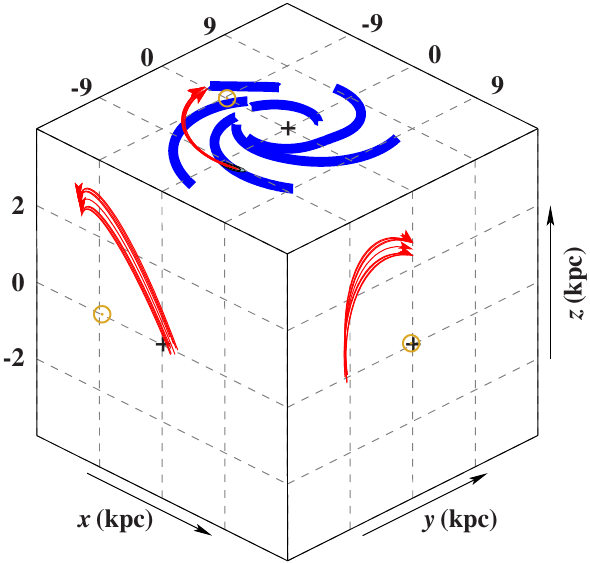}
\caption{The orbit of J0833+4308 in a Galactic Cartesian
coordinate system with the z-axis pointing to the Galactic north pole.
Red lines show nine trajectories computed with the mass model I of \citet{andreassmass} taking into account the uncertainties in distance, proper motions, and radial velocity. The arrows indicate the star’s current position. Orbits were computed back in time until they reached the Galactic
plane.  The thick blue solid lines depict the spiral arms 78 Myr ago based on the schematic model of \citet{2014A&A...569A.125H} 
and the Galactic rotation curve \citep[from model I in][]{andreassmass}. {The small shaded areas are contours for the intersection of the Galactic plane (1$\sigma$ in red and 2$\sigma$ in light blue). Note that they coincide perfectly with a section of one spiral arm (Carina-Sagittarius).}
The yellow circled
dot indicates the current position of the Sun, while the black plus sign (+) marks Galactic center. The orbit is characteristic
of a disk runaway star.}\label{fig:cube}
\end{SCfigure*}

\section{Kinematic analysis}

Proper motion components from {\it Gaia} EDR3 combined with the radial velocity
and distance from spectroscopy allows us to derive the transversal velocity. Its present Galactic rest frame velocity is $210\pm4$ km\,s$^{-1}$, much lower than the Galactic escape velocity, and therefore, the star is bound to the Galaxy. Calculations of Galactic trajectories in a gravitational potential as described by \citet{andreassmass} allow us to identify the place of birth of the star. 
Fig. \ref{fig:cube} shows the traceback of the trajectory to the Galactic disk and the Galactic disk crossing position and velocity, as well as ejection velocity and time of flight are listed in Table \ref{tab:kinematic}. The Galactic crossing footprint lies within the 
Carina-Sagittarius spiral arm, consistent with the run-away scenario from star forming regions and young open clusters.

The star was ejected 78$\pm4$ Myrs ago from the Galactic plane at a velocity of 92$\pm6$ km\,s$^{-1}$ typical for run-away B stars of similar mass in the halo \citep[see][]{2011MNRAS.411.2596S}. Such an ejection velocity is consistent with simulations of dynamical ejection from star clusters \citep[e.g.][]{2020MNRAS.495.3104S} as well as with the binary supernova ejection (BSE). The time of flight, however, is significantly shorter than its evolutionary age. This may be at odds with DES, if the star was ejected during cluster formation or soon thereafter, as predicted by simulation models \citep{2011Sci...334.1380F}.
In the binary supernova ejection (BSE) scenario, however, a delay between binary formation and ejection is expected, because the massive primary has to evolve into a supernova before the secondary is ejected as a runaway star. The difference between the age of the star $104^{+11}_{-13}$ Myr and the time of flight ($78\pm4$ Myr) is in agreement with the evolutionary time of a star more massive than 8 M$_\odot$ until core collapse.  
Therefore, we consider the BSE scenario to be the more likely one. 

We traced back the star to possible parent clusters using the method introduced in \citet{bhat} and the \textit{Gaia} DR3 based open cluster catalog of \citet{2023A&A...673A.114H}. We were unable to find any parent clusters, which could be attributed to the fact that the open cluster census is not complete for other spiral arms. 

\begin{table}
\caption{Galactic restframe velocity $v_{\textnormal{grf}}$, plane crossing positions ($x_{\textnormal{d}}$,$y_{\textnormal{d}}$,$r_{\textnormal{d}}$), plane crossing velocities ($v_{x\textnormal{,d}}$,$v_{y\textnormal{,d}}$,$v_{z\textnormal{,d}}$), disk ejection velocity $v_{\textnormal{ej}}$, and time of flight $\tau_{\textnormal{flight}}$. }\label{tab:kinematic}
\renewcommand{\arraystretch}{1.0}
\begin{center}
\begin{tabular}{lr}
\hline\hline
$v_{\textnormal{grf}}$ [km\,s$^{-1}$]   &  $210\pm4$  \\
$x_{\textnormal{d}}$ [kpc]&                                   $1.60\pm0.7$  \\
$y_{\textnormal{d}}$ [kpc]&                                   $-9.49\pm0.21$ \\
$r_{\textnormal{d}}$ [kpc]&                                   $9.63\pm0.19$ \\
$v_{x\textnormal{,d}}$ [km\,s$^{-1}$] &                   $-247\pm4$  \\
$v_{y\textnormal{,d}}$ [km\,s$^{-1}$] &                   $-54\pm16$ \\
$v_{z\textnormal{,d}}$ [km\,s$^{-1}$] &                   $90\pm6$ \\
$v_{\textnormal{ej}}$  [km\,s$^{-1}$] & $92\pm6$ \\
$\tau_{\textnormal{flight}}$[Myr] &           $78\pm4$ \\
\hline
\end{tabular}
\end{center}
\end{table}

\section{\textit{TESS} light curve}\label{sect:lc}

\textit{TESS} observed J0833+4308 (aka TIC 409272428) with 2 minute cadence in Sector 47 for 27 d with a gap of 3.5 d. 
We used the Python package \textsc{lightkurve}\footnote{\url{https://lightkurve.github.io/lightkurve}} to download the TESS data. 
The Pre-search Data Conditioning SAP flux (PDCSAP) was used, which has long term trends removed from the simple
aperture photometry (SAP), and corrects for
contributions to the aperture from
neighboring stars. Those contributions are expected because of the large pixel size  of \textit{TESS}
(almost 21 arcsec). However, this correction is not perfect and caution is necessary, if too much of the flux originates from other sources than the target. 
The estimate of how much flux in the aperture belongs to the target \citep[see also][]{2022A&A...666A.182S} is given in the CROWDSAP parameter. 
For J0833+4308 almost 98\% of the light (CROWDSAP parameter of 0.97733873) comes from the target. Hence, we can ignore the contaminating light and analyse its light curve.

We calculated a Lomb Scargle periodogram (Lomb 1976; Scargle 1982) of the light curve up to the Nyquist frequency with an oversampling by a factor of ten, using the function available in the \textsc{lightkurve} package. The periodogram shows three neighbouring peaks around $3.7747\pm0.0041$~d ($f=0.26492\pm0.00029$~1/d). However, the window function, which we calculated using the code provided by Keaton Bell\footnote{\url{https://gist.github.com/keatonb/51b52a7d564e6b470421f4b5b8cad4ed\#file-spectralwindow-ipynb}} shows that the smaller peaks are created by the limited data, which is only a few periods long.

\begin{figure*}
\begin{center}
\includegraphics[width=0.58\textwidth]{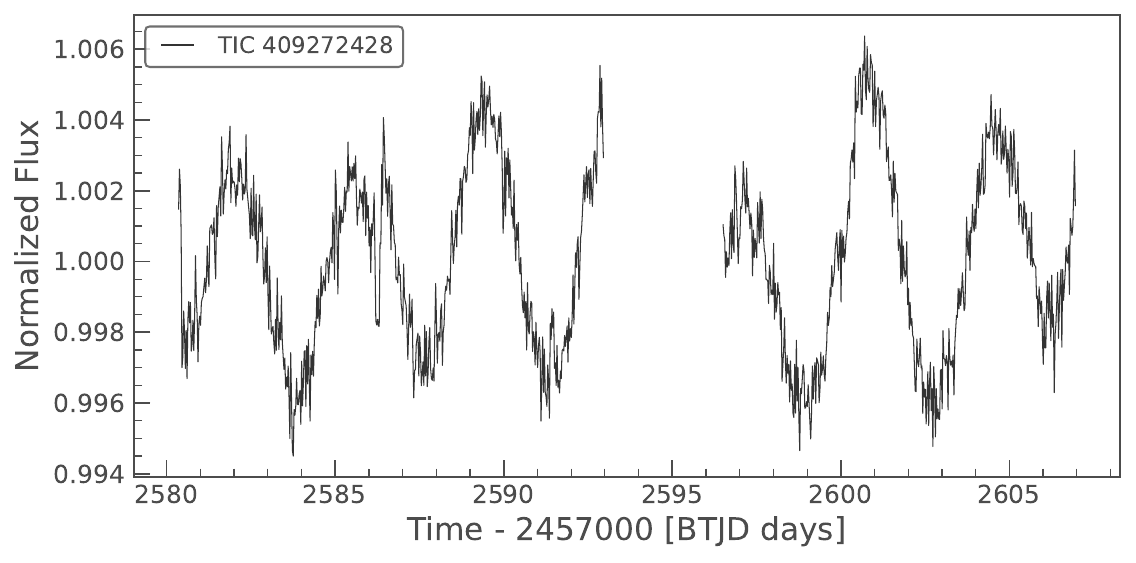}
\includegraphics[width=0.41\textwidth]{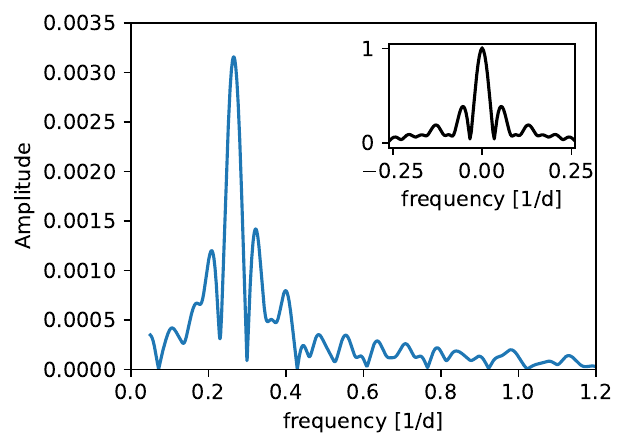}
\end{center}
\caption{\textit{TESS} light curve of J0833+4308 (left) and its Fourier transform (right) including the window function in the inset.}\label{fig:lc_fc}
\end{figure*}

The periodic variation (see Fig. \ref{fig:lc_fc}) with 3.77 days period has an amplitude of 3~mmag, which is typical for Slowly Pulsating B stars (SPBs). The amplitude is varying, indicating beating, which is caused by additional periodicities close to the main peak, which cannot be resolved due to the short light curve. To look for additional periods we used the python package \textsc{pywhiten}\footnote{https://pywhiten.readthedocs.io/}. No other significant peaks could be found. However, there are discrepancies between the observed periodogram and the window function, which could indicate nearby close peaks. 

B stars also have been found to show light variations with the rotational period due to spots (Shen et al. 2023). As we have only an upper limit for $v\sin i$ and {available data are insufficient to} constrain the inclination, we do not know the rotational period. Typically, they are in the range of a few days. Due to the differential rotation, often alias peaks with half the orbital periods and additional periodicities near the main period are observed. We do not see any alias peaks.
Accordingly, the light variations of J0833+4308 are likely caused by stellar pulsations as its main oscillation period is typical for slowly pulsating B-stars. However, we cannot exclude that the variations are due to spots. More time-resolved photometry would be needed to distinguish between spots and pulsations as pulsations are stable only for a few months usually.%

To our knowledge pulsations of runaway B stars have been reported in a few cases, only. PG 1610$+$062 is known to be a slowly pulsating B run-away star \citep{irrgang19} and the hotter run-away B-star PHL\,346 (TIC 69925250) to be a $\beta$ Cep pulsator \citep{waelkens88,handler19}. Hence, J0833+4308 is a rare case of most likely being a slowly pulsating runaway B star, which deserves additional photometric monitoring.

\section{Summary and outlook}

A comprehensive analysis of J0833+4308 showed it to be a slowly pulsating 4.6\,M$_\odot$ B-type main sequence star {with a chemical} composition {typical for young massive stars \citep{2012A&A...539A.143N}} in the Galactic halo ejected at a velocity of 92$\pm6$ km\,s$^{-1}$ from its place of birth, likely an open cluster located in the  
Carina-Sagittarius spiral arm about 78 My ago. Its age of 104 Myrs is somewhat larger than the time of flight, which favours the binary supernova   
most likely through the binary supernova scenario. Accordingly, the star was ejected when  a binary system was disrupted by the supernova explosion of the massive primary. The remnant of the primary is a neutron star or black hole. Run-away neutron stars linked with B run-away stars have been found previously, e.g. the runaway neutron star PSR B1706-16 which has been shown to be associated with $\zeta$ Oph \citep[][]{2020MNRAS.498..899N,bhat}.

 The spectroscopic results presented here
 are limited by the low resolution of available spectra. High resolution spectroscopy is required to improve the accuracy of the chemical abundances, extend the abundance analyses to other chemical elements, determine the projected rotation velocity, and to quantify line profile variability from pulsations. To resolve the pulsation frequency spectrum additional \textit{TESS} observations are necessary. A search for an associated runaway neutron star would be challenging due to the long flight time.


\section{Acknowledgements}
A.B. was supported by the Deutsche Forschungsgemeinschaft (DFG) through grant GE2506/18-1.
We thank Andreas Irrgang and Matti Dorsch for developing and maintaining their analysis tools and making them available to us.
This work has made use of data from the European Space Agency (ESA) mission {\it Gaia} (\url{https://www.cosmos.esa.int/gaia}), processed by the {\it Gaia} Data Processing and Analysis Consortium (DPAC, \url{https://www.cosmos.esa.int/web/gaia/dpac/consortium}). Funding for the DPAC has been provided by national institutions, in particular the institutions participating in the {\it Gaia} Multilateral Agreement.
This research has made use of NASA's Astrophysics Data System. This paper includes data collected by the TESS mission. Funding for the TESS mission is provided by the NASA's Science Mission Directorate.

\bibliography{runaway_star.bib}
\end{document}